\documentclass[twocolumn,showpacs,pra]{revtex4-1}
\usepackage{graphicx}

\usepackage{amsmath, amssymb, color}

\def\a{\alpha}

\def\om{\omega}
\def\Om{\Omega}

\def\ve{\varepsilon}

\definecolor{grey}{rgb}{0.4,0.4,0.5}
\definecolor{darkgreen}{rgb}{0,0.5,0}
\definecolor{darkred}{rgb}{0.6,0.0,0}
\definecolor{lightbrown}{rgb}{1,0.9,0.8}
\definecolor{brown}{rgb}{0.6,0.3,0.3}
\definecolor{darkblue}{rgb}{0,0,0.8}
\definecolor{darkmagenta}{rgb}{0.5,0,0.5}

\newcommand{\pDer}[2]{{{\partial} #1 \ov {\partial} #2}}

\def\({\left(}
\def\){\right)}

\def\be{\begin{equation}}
\def\ee{\end{equation}}
\def\ben{\begin{equation*}}
\def\een{\end{equation*}}

\newcommand{\bea}{\be\begin{aligned}}
\newcommand{\eea}{\end{aligned}\ee}
\newcommand{\bean}{\ben\begin{aligned}}
\newcommand{\eean}{\end{aligned}\een}

\newcommand{\bei}{\begin{itemize}}
\newcommand{\eei}{\end{itemize}}

\newcommand{\bee}{\begin{enumerate}}
\newcommand{\eee}{\end{enumerate}}

\newcommand{\bem}{\left (\begin{matrix}}
\newcommand{\eem}{\end{matrix} \right )}

\def\ov{\over}

\newcommand{\sfrac}[2]{{\textstyle\frac{#1}{#2}}}
\newcommand{\half}{\sfrac{1}{2}}

\def\hb{\hbar}

\def\det{\hbox{det}}
\def\mA{\mathcal A}

\begin{document}
 
\title{Modulated trapping of interacting  bosons in one dimension}

\author{Eoin Quinn}
\author{Masudul Haque}

\affiliation{Max-Planck-Institut f\"ur Physik komplexer Systeme, N\"othnitzer Str. 38, 01187
  Dresden, Germany} 

\date{\today}

\begin{abstract}

  We investigate the response of harmonically confined bosons with contact interactions (trapped
  Lieb-Liniger gas) to modulations of the trapping strength.  We explain the structure of resonances
  at a series of driving frequencies, where size oscillations and energy grow exponentially.  For
  strong interactions (Tonks-Girardeau gas), we show the effect of resonant driving on the bosonic
  momentum distribution.  The treatment is `exact' for zero and infinite interactions, where the
  dynamics is captured by a single-variable ordinary differential equation.  For finite interactions
  the system is no longer exactly solvable.  For weak interactions, we show how interactions modify
  the resonant behavior for weak and strong driving, using a variational approximation which
  adds interactions to the single-variable description in a controlled way.

\end{abstract}



\maketitle 


\section{Introduction}\label{sec:intro}

Periodic driving of quantum systems has been of interest for many decades, since the
early period of quantum theory \cite{early_periodic_driving_papers}.  
The question of how a quantum system evolves after a time-periodic perturbation has been turned on
is natural in various contexts, e.g. as a model of subjecting quantum matter to electromagnetic
radiation.  In textbooks, this is often discussed in the context of using time-dependent
perturbation theory to calculate transition rates (e.g. Fermi's golden rule).  Periodic driving and
its treatment using the Floquet picture has also been an important paradigm in the analysis of
nuclear magnetic resonance (NMR) \cite{NMR_driving}.

More recently, experimental developments with ultracold trapped atoms has revived interest in this
paradigm, in particular for many-body quantum systems.  For this class of experiments, it is
possible to control, and in particular periodically modulate, many different parameters.  
\emph{Modulation spectroscopy} has become a standard tool in investigating the excitation spectrum
of many-body systems \cite{periodic_expts_other, periodic_expt_Arimondo_Fazio, Naegerl_Science2009}.
Modulating the strength of an optical lattice at various frequencies, the energy absorbed by the
system is found to be maximal when the frequency matches possible excitation energies of the
many-body system.
%


Since most cold-atom experiments involve a near-harmonic trapping potential, parameters relating to
the trap are of special interest.  In the experiment and calculations of Ref.\
\cite{periodic_expt_Arimondo_Fazio}, modulation spectroscopy on a lattice system was performed by
driving of the trapping strength, as opposed to the more usual driving of the lattice depth.  The
effect of modulating the position of the trap center on a harmonically trapped single particle has
been widely studied; in the Floquet description, this situation is exactly solvable
\cite{position_driving}.  The modulation of the strength of the trapping potential for a single
particle has also been considered \cite{trapstrengthdriving_previous}.  The single-particle harmonic
oscillator has a unusual response to modulation due to its spectrum being equally spaced: when the
ground state is at resonance with one excited state, this excited state is at resonance with a
further excited state, and so on.  Resonant driving in this ideal system induces the energy to grow
without bound.  In the case of trap strength modulation, the energy grows \emph{exponentially} with
time.

In this work we consider one-dimensional boson systems with contact interactions (Lieb-Liniger gas
\cite{LiebLiniger1963}), subject to a perfectly harmonic trap.  Considering the modulation of the strength
of such a trap, we describe the resulting dynamics of the bosonic system, which is initially in the
ground state of the un-modulated Hamiltonian.  We clarify the energy absorption response and the
dynamics of the cloud size (radius) at different frequencies.

The Hamiltonian is
\begin{multline}
H =  {1 \over 2m} \sum_{j=1}^N \Big[ -\hb^2\pDer{^2}{x_j^2} ~+~ m^2\om^2(t) \hat x_j^2\Big] \\ 
~+~ U \sum_{1\leq j< k \leq N} \delta(\hat x_j - \hat x_k) \,, 
\end{multline}
where $N$ is the number of bosons in the trap, $U$ is the interaction strength, $m$ is the boson
mass.

We will examine the time evolution under periodic modulation of the trapping strength, so that
$\om(t+T)=\om(t)$.  We will present results for modulation of the form
\begin{equation}
\om^2(t) = \begin{cases} \om^2_0 & t< 0 \,,\\ 
 \om^2_0(1+ \lambda \sin \Om t) & t>0 \,.\end{cases} 
\end{equation}
The system is initially taken to be the ground state of the $\omega=\omega_0$ Hamiltonian.  We
examine how this system evolves in time once the modulation is turned on after time $t=0$.

Ultracold bosonic atoms behave as a 1D system when the transverse degrees of freedom are frozen out
by tight confinement \cite{Olshanii_PRL1998, Petrovetal_PRL2000}.  The system of Lieb-Liniger bosons
in a harmonic trap has by now been realized in multiple cold-atom labs \cite{Naegerl_Science2009,
  trapped_1D-boson_expts}.  Controlled modulation of the trap strength is also standard
\cite{periodic_expt_Arimondo_Fazio}.  Therefore, the questions addressed in the present work can in
principle be explored experimentally in one of several laboratories.  In practice, deviations from
exact harmonicity might spoil the effect of exponential resonances, since such deviations destroy
the equal spacing of the spectrum.  

For a single particle in a driven harmonic trap, the dynamics can be described by a single-variable
ordinary differential equation for the spatial size of the wavefunction.  At zero interaction $U=0$,
the single-particle description is sufficient to describe the ideal condensate dynamics.  In Section
\ref{sec:nonint}, we describe and characterize the resonance structure for a single particle and
(equivalently) for the non-interacting gas.  In Section \ref{sec:TG}, we treat the case of very
strong coupling (Tonks-Girardeau limit).  In the $U\to\infty$ limit the many-body bosonic system can be
mapped to non-interacting fermions \cite{Girardeau60-65}, and  the dynamics can then be constructed from
the dynamics of single particles which start at different single-particle eigenstates.  We also
present the behavior of `off-diagonal' or `bosonic' properties which distinguish the Tonks-Girardeau
gas from free fermions, namely the momentum distribution and the natural orbital occupancies.  The
momentum distribution $n(p)$ is simple for non-resonant cases and has typical bosonic form, but when
driven resonantly, it undergoes periodic `fermionization'.  We also show that the natural orbital
occupancies have the peculiarity that they do not change at all for arbitrary time-variations of the trap
strength.  Finally in Section \ref{sec:GP}, we treat finite nonzero $U$.  In this case an exact
treatment is beyond reach.  We use a mean field treatment together with a time-dependent variational
ansatz that provides a single-parameter description of the dynamics.  The parameter is again the
size of the cloud, which gives a satisfying way to compare to the exactly solvable $U=0$ case.  We
show how the exponential resonances are killed at finite $U$ for weak driving, but survive for
stronger driving.



In the figures, we will plot all quantities in `trap units' appropriate to the initial or mean
frequency $\omega_0$, i.e. energy, time, distance and momentum are expressed in units of 
\begin{equation} \label{eq:trapunits}
\hbar\omega_0 
\,,\quad  \frac{1}{\omega_0}
\,,\quad  \sqrt{\frac{\hbar}{m\omega_0}}
\,,\quad  \sqrt{\hbar{}m\omega_0}
\end{equation}
respectively.

\begin{figure}[tb]
\centering
\includegraphics[width=0.999\columnwidth]{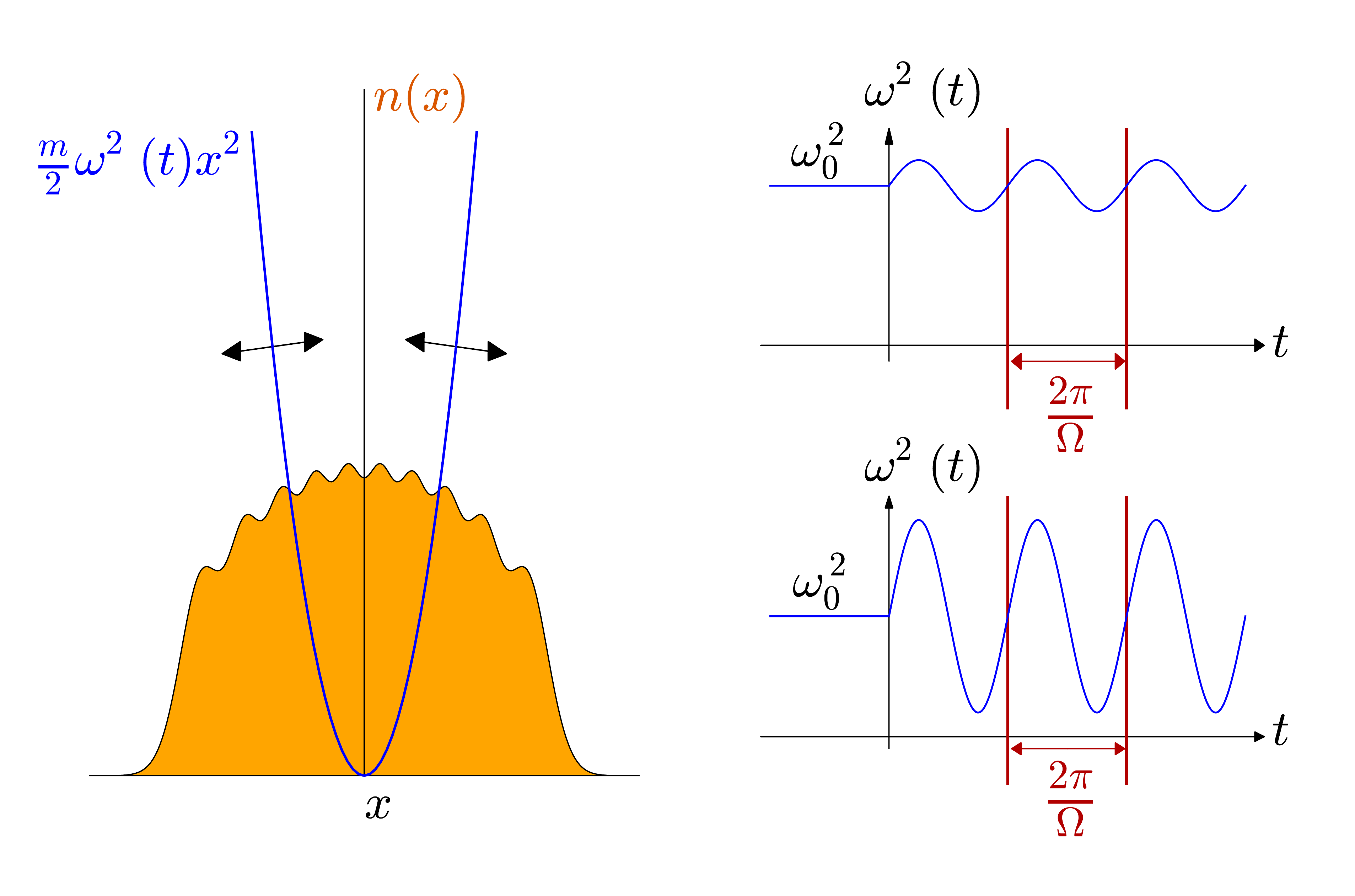}
\caption{\label{fig:trap}
Left: density profile of a (strongly interacting) Bose gas in a harmonic trap.  At $t=0$ the trapping
strength begins to oscillate, as indicated by the arrows.  Right: definition of driving frequency
$\Omega$. A weak ($\approx$ perturbative) and a strong driving case are shown.  
}
\end{figure}

\section{Non-interacting Bosons ($U=0$)  \label{sec:nonint}}   

The simplest case is that of non-interacting bosons, $U=0$.  The problem then reduces to that of a
single particle in a driven quantum harmonic oscillator, since the initial state is an ideal Bose
condensate with all particles in the lowest harmonic oscillator state.  

The driven single-particle problem has been discussed in the literature, in particular in the
presence of dissipation \cite{trapstrengthdriving_previous}.  Methods of solution for this problem,
based on a scaling transformation or on Floquet theory, are well-known.  For completeness, in this
section we describe briefly the scaling transformation that provides the solution for
\emph{arbitrary} temporal variations of the trapping strength, in terms of a single ordinary
differential equation.  We also describe the resonance structure for both small and large driving.

\subsection{Exact solution \label{sec:SP}}   

If the initial state is an eigenstate of the initial harmonic oscillator Hamiltonian, then the time
evolution under time-dependent trapping strength is given by the scaling form
\cite{KaganSurkovShlyapnikov_PRA96, Perelomov_book_QM, GritsevDemler_NJP2010, MinguzziGangardt2005,
  GangardtPustilnik_PRA2008}
\begin{equation}\label{eq:phit}
\psi_n(x,t) = {1\ov \sqrt b}\, \exp\Big[ {im\ov
  2 \hb}{\dot b \ov b}x^2 - {i\ov \hb} \ve_n t \Big]\,\psi_n\big({x\ov b},0\big) \,,   
\end{equation}
where $ \psi_n(x,0)$ is an eigenstate of the initial Hamiltonian, with eigenenergy
$(n+\half)\hbar\omega_0$.  The wavefunction changes scale and the phase evolves, but the shape
remains unchanged, for arbitrary time dependence $\omega_0(t)$ of the trapping strength.  For
example, if starting in the ground state, the wavefunction magnitude retains Gaussian form.  
The scale factor $b(t)$ contains all the information about time evolution.  It obeys the
second-order differential equation
\begin{equation}\label{eq:b}
\ddot b + \om^2(t)b = {\om_0^2 \ov b^3}\,,
\end{equation}
with initial conditions $b(0)=1$ and $\dot b(0)=0$.  
It determines the quasienergies through
\be
\ve_n(t) = {1\ov t} \int_0^t\, dt'\,{(n+\half) \hb \om_0 \ov b^2(t') } \,.
\ee

The energy at time $t$ is given in terms of $b(t)$ and its time
derivative: 
\begin{equation}
E_n(t) =(n+\half) \hb \om_0 \Big[ {\dot b^2\ov 2\om_0^2} + {\om^2(t) b^2 \ov 2 \om_0^2} +{1\ov 2b^2} \Big] \,.
\end{equation}
For a system of  $N$ non-interacting bosons, all the bosons start at the $n=0$ single-particle state
and the evolution of each is described by the above equations.  The energy of the system is  given by
\begin{equation} \label{eq:EU0}
E_{U=0}(t) = N E_0(t)\,.
\end{equation}

\subsection{Structure of resonances}   

The resonance structure for the single particle has been touched upon previously in the literature,
at least for related situations, e.g. in \cite{trapstrengthdriving_previous}.  However we have not
seen a complete description, especially for strong driving; so we provide one in this section.

We first describe small-amplitude  driving.

Since we are driving the trap strength $\half{m}\omega^2x^2$, and $\hat x^2$ connects wave functions
that differ in energy by $2\om_0$, the relevant energy gap is $2\om_0$.  The structure of resonances
is that obtained by thinking about the ground state and first even-parity excited state ($n=0$ and
$n=2$ states) as forming a two-level system with energy separation $2\omega_0$.  As in a two-level
system where both diagonal and off-diagonal terms are driven sinusoidal \cite{twolevel_system},
there are resonances at all frequencies $\Omega=2\omega_0/j$, where $j$ is a positive integer.  The
primary resonance at $\Omega=2\omega_0$ ($j=1$) directly excites the first even-parity excited
state.  The higher-order resonances ($j>1$) correspond to multi-photon processes where the primary
gap is excited by multiple quanta of energy.

These are also the only resonances; for example, there are no resonances at
$\Omega=2\omega_0{\times}j$ with $j>1$, even though such frequencies match the energy difference
between the ground state and higher even-parity states.  The reason is that the $x^2$ operator does
not connect such pairs of eigenstates. 


\begin{figure}[tb]
\centering
\includegraphics[width=0.99\columnwidth]{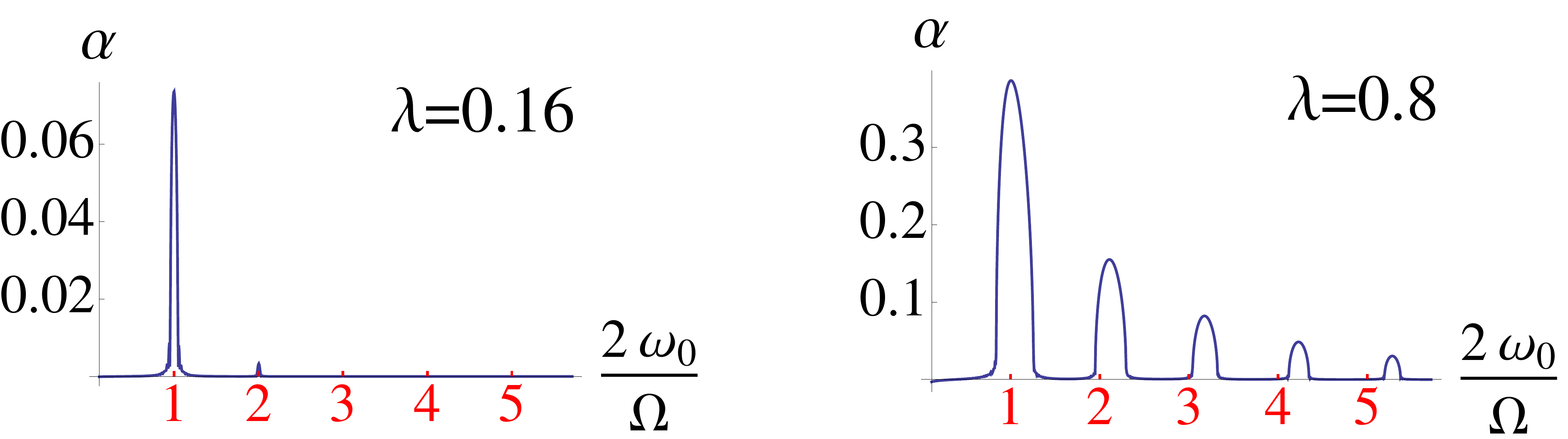}
\caption{\label{fig:RC}
Resonance curves for two values of $\lambda$. The quantity $\a$ is defined in 
Eqs.\ \eqref{eq:alpha}, \eqref{eq:alpha1}.  For weak driving, the primary resonance is at
$\frac{2\omega_0}{\Omega}=1$, the sub-resonances (`multi-photon' resonances) are at higher integer values of
$\frac{2\omega_0}{\Omega}$.  With increasing driving
strength the sub-resonances grow in magnitude, the resonance peaks broaden, and the positions of
the resonances drift. 
}
\end{figure}

For small driving strength $\lambda$, there are sharp resonances exactly at $\Omega=2\omega_0/j$.
For large $\lambda$, the resonances get broadened and also shifted.  To  illustrate
these features, we present `resonance curves' in Figure \ref{fig:RC} for $\lambda=0.16$ and
$\lambda=0.8$.   

In characterizing the resonances, one has to take into account the exponential nature of the
resonances.  Although the resonance positions can be understood by regarding the first two
even-parity states as forming a two-level system, the physical consequence of resonant driving is
more drastic because of the unbounded and equally spaced spectrum: resonant population of the $n=2$
state leads to successive resonant population of the higher even-$n$ states, so that the energy
increases exponentially at resonance.  To quantify the long time energy absorption we define
\begin{equation}\label{eq:alpha}
\a(\Om) = \Big\langle \frac{\log \big(E_0(t)/E_0(0)\big)}{t} \Big\rangle_{t\to\infty} 
\end{equation}
where $\langle\ldots\rangle_{t\to\infty}$ denotes a time average taken in the long time limit.  This
quantity is zero when there is no exponential increase of energy.  For the purposes of Figure
\ref{fig:RC}, $\a$ is evaluated approximately as
\begin{equation}  \label{eq:alpha1}
\a(\Om) = {1\ov N_p}\sum_{j=1}^{N_p} {1\ov t_j} \log\big(E_0(t_j)/E_0(0)\big)\,,
\end{equation}
with $N_p=40$ and $t_j=30{2\pi\ov \Omega} + j {\pi\ov\Omega}$, i.e. the average is taken over the
ten periods starting from the thirtieth using four points in each period.  Some dependence on $N_p$
and the discretization $t_j$ is expected, but we believe that the curves shown in Figure
\ref{fig:RC} are converged sufficiently for the principal features to be clearly seen.

In Figure \ref{fig:RC}, to clearly display the resonant peaks, $\a$ is plotted against
$2\omega_0/\Omega$, so that the primary resonance occurs at 1, and the sub-resonances occur at
higher integer values.  Resonances beyond the primary resonance are suppressed; this effect is
stronger for small $\lambda$.  For large $\lambda$ the locations of the sub-resonances are
significantly shifted in addition to the expected broadening.

\begin{figure}[tb]
\includegraphics[width=0.99\columnwidth]{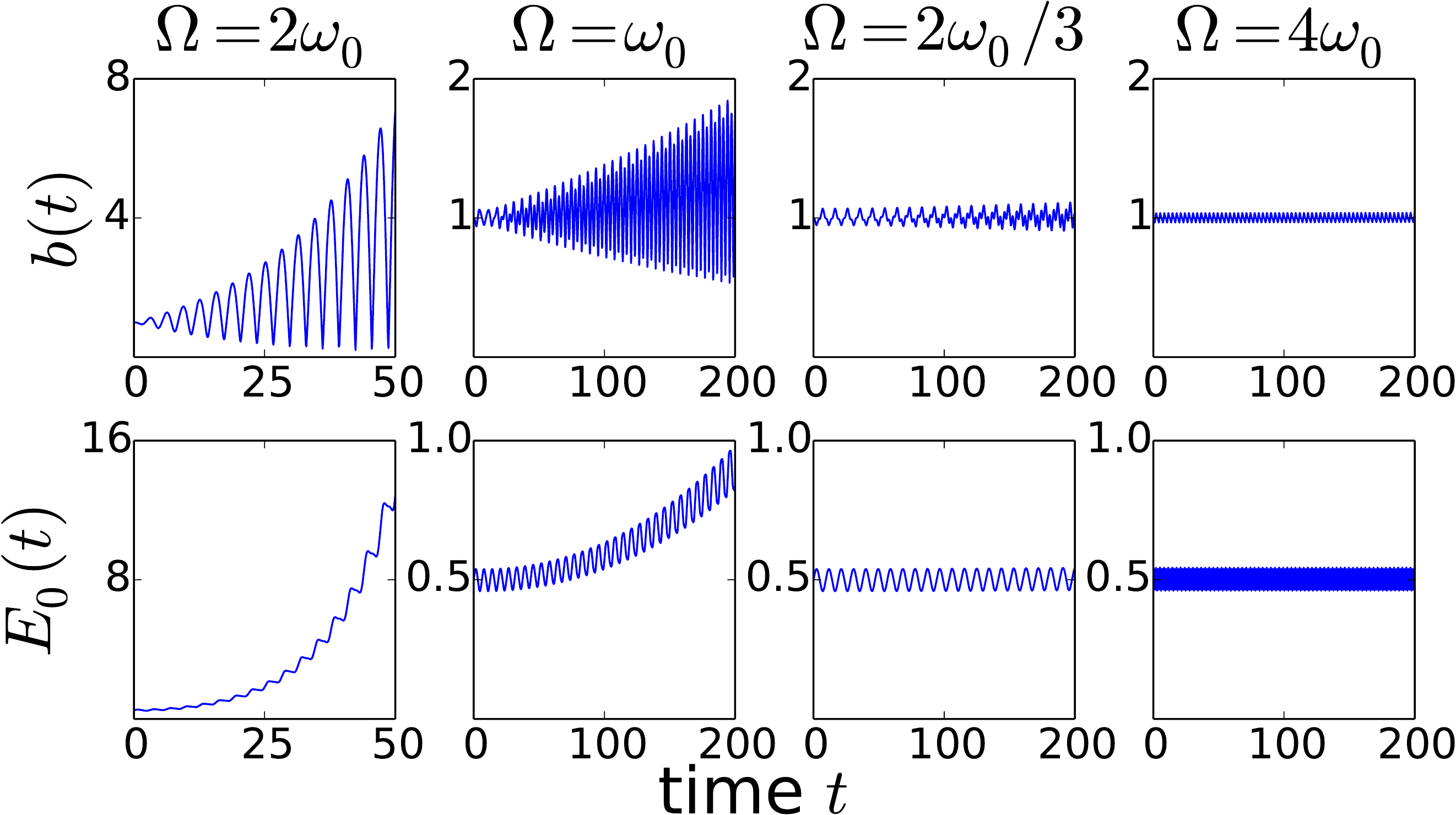}
\caption{ \label{fig:bE1}
The scaling function $b(t)$ and the energy $E_0(t)$ for several driving frequencies $\Om$, with relatively
weak driving, $\lambda =0.16$.  There are resonances at the first three of the shown frequencies, The
resonances at $\Omega=2\omega_0/j$ get weaker for increasing $j$.  For $\Omega=2\omega_0/3$, the
exponenital increase is barely visible at the time scales shown. Time and energy are expressed in
trap units, Eq.\ \eqref{eq:trapunits}.
}
\end{figure}

\begin{figure}[tb]
\centering
\includegraphics[width=\columnwidth]{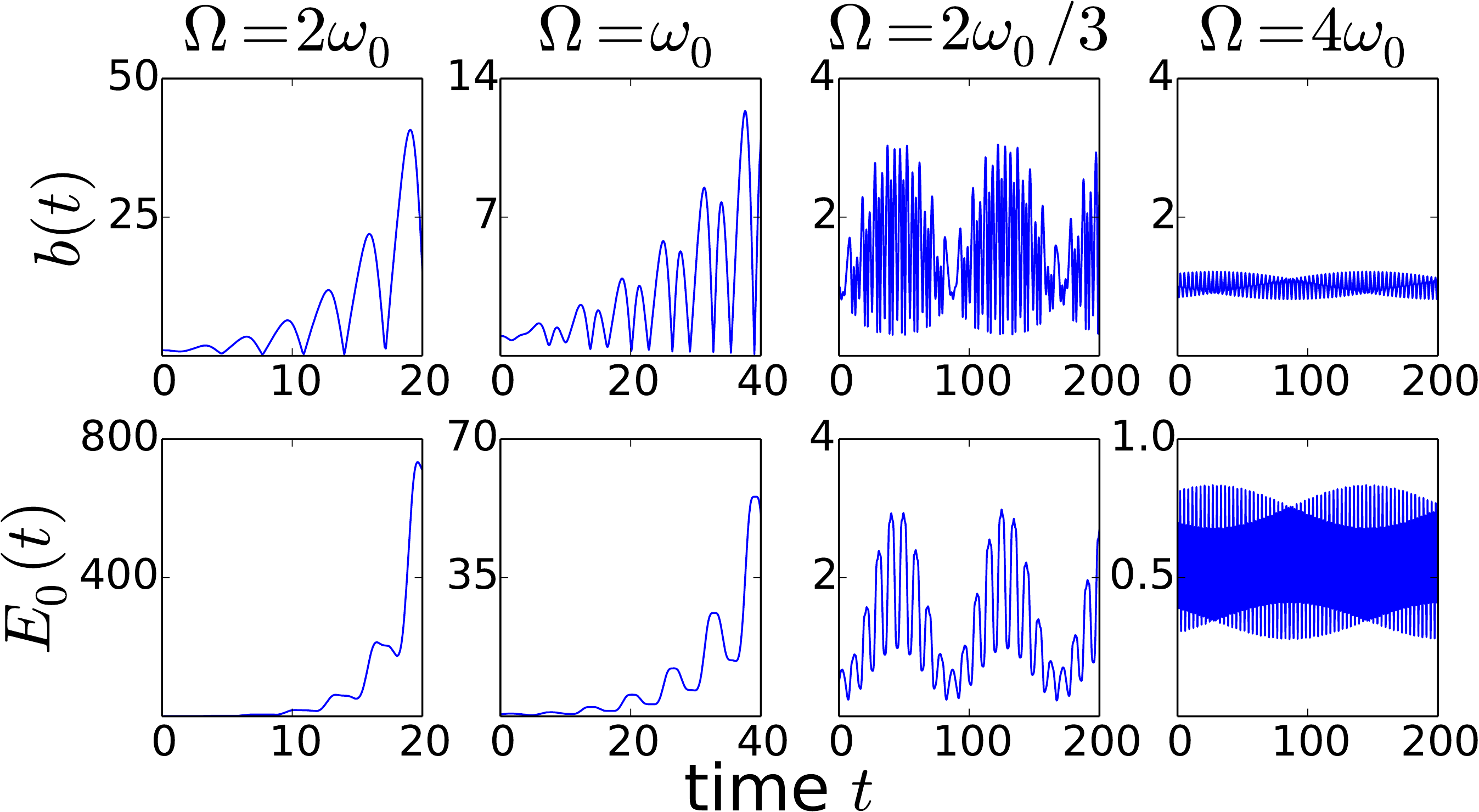}
\caption{\label{fig:bE2}
The scaling function $b(t)$ and the energy $E_0(t)$ for several driving frequencies $\Om$, with
strong driving, $\lambda =0.8$.  The resonance at
$\Omega=2\omega_0/3$ is now lost due to the shift of the resonance.   Time and energy are expressed in
trap units, Eq.\ \eqref{eq:trapunits}.
}
\end{figure}

In figures \ref{fig:bE1} and \ref{fig:bE2} the time evolution of both the scaling function $b(t)$
(size of the wavefunction) and the energy $E_0(t)$ of a single particle are plotted for various
driving frequencies at the two different couplings, $\lambda=0.16$ and $\lambda=0.8$.  The suppression of
sub-resonances for small $\lambda$ and shifting of the resonant peaks for large $\lambda$ (both effects
discussed above and seen in Figure \ref{fig:RC}) can also be seen in these time evolution plots.  In
particular, the third resonance is shifted away from $\Omega=2\omega_0/3$ at large $\lambda$, so
that no exponential increase is seen at this driving frequency.

\section{Strong Interactions  ($U\to\infty$): The Tonks-Girardeau gas   \label{sec:TG}}   

The TG gas has an exact solution consisting of two parts. Firstly the model is mapped onto free
fermions in a trap, and secondly the time evolution of this system is solved by the scaling
transformation described in the previous section. The mapping to fermions is achieved by an
anti-symmetrisation of the wave function \cite{Girardeau60-65}
\begin{equation}
 \Psi_{TG}(x_1,\ldots,x_N;t) = \mA(x_1,\ldots,x_N)  \Psi_{F}(x_1,\ldots,x_N;t) \,,
\end{equation}
with the factor $\mA(x_1,\ldots,x_N) = \prod_{1\leq j< k \leq N} \mbox{sgn} (x_j-x_k)$.  

Since we start from the ground state, the initial fermonic wave function involves the Slater
determinant of the the $N$ particles placed in the first $N$ harmonic oscillator eigenfunctions,
$\psi_j(x,0)$.  Under driving, the time evolution of these single-particle orbitals $\psi_N(x,t)$
can be obtained from Eqs.\ \eqref{eq:phit} and \eqref{eq:b}.  The time-dependent many-fermion
wavefuction is the Slater determinant formed with these time-dependent single particle orbitals:
\begin{equation}
\Psi_{F}(x_1,\ldots,x_N;t) = {1\ov \sqrt{ N!}} \det_{j,k=1}^N \psi_j(x_k,t)\, .
\end{equation}

Thus the time evolution for the wave function of the $N$-particle TG gas, starting from the ground
state at $t=0$, is given by
\begin{multline}
\Psi_{TG}(x_1,\ldots,x_N;t) = \\ b^{-N/2} \Psi_{TG}(x_1/b,\ldots,x_N/b;0) \times \\ \exp\Big({im\ov 2 \hb}{ \dot b \ov b } \sum_j x_j^2  \Big) \exp(-{i\ov\hb} N^2{ \ve_0}t\big)\,.
\end{multline}
and the corresponding energy of the system is
\begin{equation}\label{eq:ETG}
E_{TG}(t) = \sum_{n=0}^{N-1}\big(2 n+1\big) \;E_0(t)   =  N^2 E_0(t)\,.
\end{equation}

\subsection{One-body density matrix and momentum distribution} \label{sec:MP}

The one-body density matrix is given by 
\begin{multline} \label{eq:TG_densmatr}
g_{TG}(x,y;t) = N\int dx_2,\ldots,dx_N\,{\mathfrak a}(x) {\mathfrak a}(y) \times\\
~~~\Psi_F^*(x,x_2,\ldots,x_N;t)\Psi_F(y,x_2,\ldots,x_N;t)\,.
\end{multline}
where ${\mathfrak a}(x) = \prod_{j=2}^L \mbox{sgn}(x-x_j)$.  At time $t=0$ the system is prepared in
its ground state, and both $g_{TG}(x,y;0)$ can be written explicitly as Hankel 
determinants \cite{Forrester2003}. Due to the
scaling form of the time-dependent wavefunction, the time evolution of the density matrix is also
captured through a scaling transformation:
\begin{equation}\label{eq:gt}
g_{TG}(x,y;t) = {1\ov b}\; \exp\left[-\frac{im}{2\hbar}\frac{\dot{b}}{b}(x^2-y^2)\right] \;
g_{TG}\big({x\ov b},{y\ov b};0\big) \,,
\end{equation}

The diagonal part of $g_{TG}$ is the particle density.  The density profile
\begin{equation}
\rho_{TG}(x;t) = g_{TG}(x,x;t)= {1\ov b}\rho_{TG}\big({x\ov b};0\big)
\end{equation}
undergoes dynamics governed by the scaling factor $b(t)$.  The form of the many-body density stays
unchanged from the equilibrium density profile, only the scale is modified, just as in the case of a
driven single particle starting from any harmonic-oscillator eigenstate.  The shape of the
equilibrium density profile for the Tonks-Girardeau gas (or free fermions) in a harmonic trap is
relatively well-known (e.g.\ Figure \ref{fig:trap} above, Refs.\ \cite{TG_density_profile}).

The interesting correlations of the TG gas appear in the off-diagonal part of $g_{TG}(x,y;t)$, which
are captured by the momentum distribution 
\bea\label{eq:mp}
n(p,t) &={1\ov { 2 \pi}} \int dx\,dy\, g_{TG}(x,y;t) e^{ip(x-y)/\hb}\\
& = {b\ov { 2 \pi}} \int dx\,dy\, g_{TG}(x,y;0)\times\\
& ~~~ ~~~ \exp{\Big[ - i b\Big(m \dot b {x^2-y^2\ov 2\hb} - {p(x-y)\ov \hb}\Big)\Big]}\,,
\eea
where in going to the second line the integration variables were rescaled by $b$.

\begin{figure}[tb]
\centering
\includegraphics[width=\columnwidth]{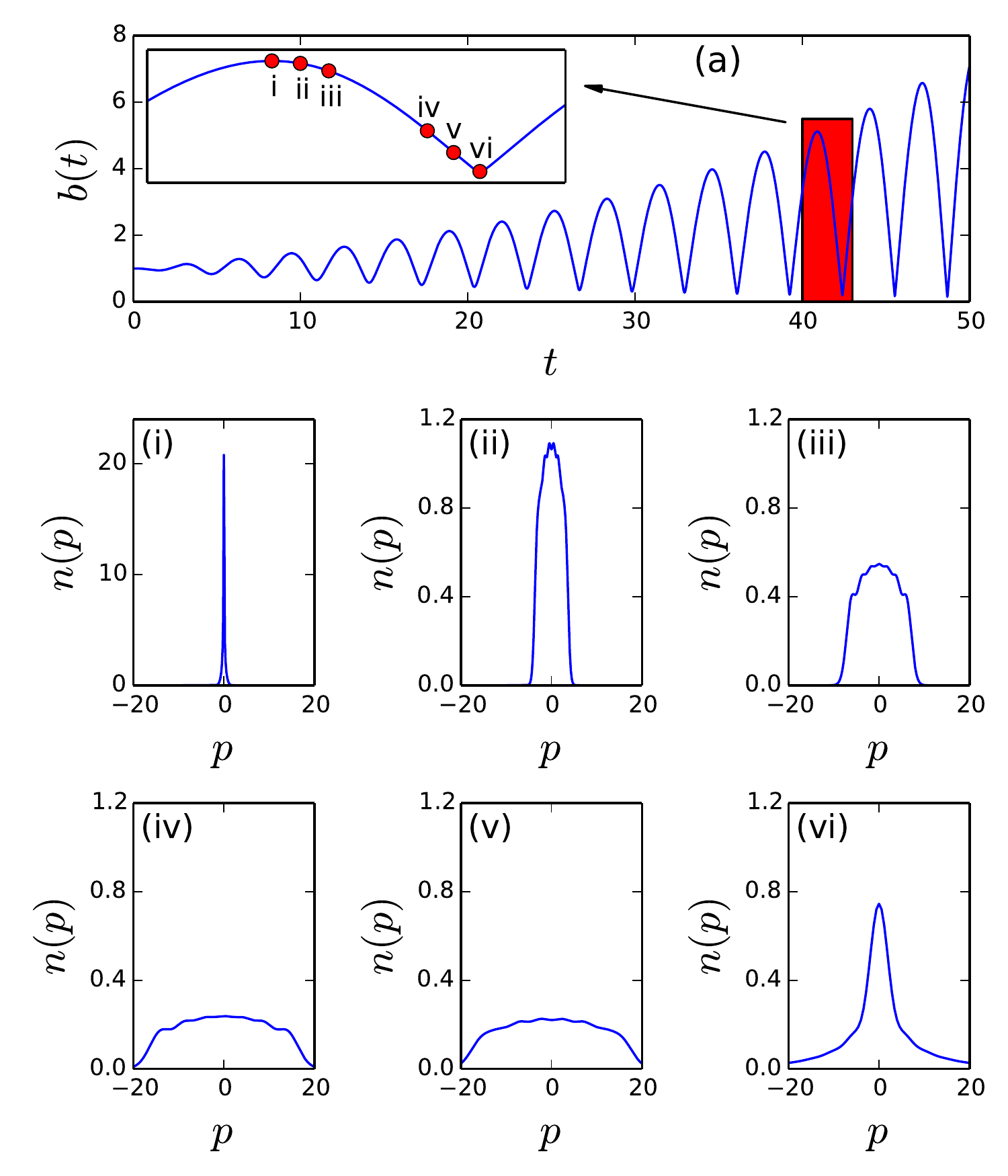}
\caption{\label{fig:np} Snapshots of the momentum distribution $n(p)$ at resonance, Tonks-Girardeau
  regime.  We show results for $N=7$ bosons driven with amplitude $\lambda=0.16$ at the primary resonance
  frequency, $\Om=2\omega_0$.  Six instants are chosen, spanning one half-period.  Note that the verticial axis
  units are different in the first snapshot.   Time and momentum are expressed in
trap units, Eq.\ \eqref{eq:trapunits}.
}
\end{figure}

The dynamics of the momentum distribution following a trap release was studied in
Refs.~\cite{MinguzziGangardt2005, GangardtPustilnik_PRA2008}. There it was found that, when the
scaling parameter $b$ is large, the momentum distribution looks like the free-fermion momentum
distribution, which is an $N$-peak distribution of the same form as the density distribution.  A
richer version of this ``dynamical fermionization'' occurs in our driven case, at resonance.  At the
exponential resonances, the scaling parameter $b$ exhibits oscillations with exponentially growing
amplitude.  As long as the derivative $\dot{b}$ is not too small, the same argument holds: making a
stationary phase approximation, the dominant contribution to the integral \eqref{eq:mp} comes from
the diagonal point $x^*=y^* = p/(m\dot{b})$, and  the momentum distribution approaches a rescaling of
the fermionic momentum distribution.  However, as $\dot b\to0$ the point of the integrand on which
the stationary phase approximation focuses has vanishing weight, which nullifies the fermionization
effect at the turning points of the oscillation.  Instead when $\dot{b}=0$, the contribution of the
dynamical phase to the momentum distribution vanishes, and the momentum distribution is a rescaling
of an equilibrium TG momentum distribution.  Dynamical fermionization thus appears and disappears
recurrently when the driving gives rise to an exponential resonance.   

This type of periodic fermionization can also be generated by a strong quench between two trapping
frequencies, as discussed in Ref.\ \cite{MinguzziGangardt2005}.  In the driven case, due to the
exponential nature of the resonances, even a weak resonant driving will eventually increase the
oscillations of $b(t)$ to the regime of periodic dynamical fermionization.

A picture of the time evolution of the momentum distribution is presented Figure \ref{fig:np}. In
Figure 5(a) the time evolution of the size parameter $b(t)$ is plotted for $\lambda=0.16$ at the primary
resonance $\Om=2\om_0$, and an inset focuses on a period at late times and indicates time slices at
which the momentum distribution is plotted. At (i) we have $\dot b=0$ and $b$ large and the
distribution takes the form of an equilibrium TG momentum distribution. With evolving time $|\dot
b|$ increases and the distribution develops fermionic correlations, and the $N$-particle peaks
emerge. By (iii) the principal features of the fermionic distribution have appeared, and these remain
present for some time, for which the evolution of the distribution is primarily through a
rescaling. From (iv) the fermionic correlations fade to the point (vi) where the scaling factor
turns and the momentum distribution again takes the bosonic form.

Away from resonance the scaling factor does not become large enough for fermionic correlations to
develop. Instead the momentum distribution remains primarily bosonic in nature, with some relatively
minor deviations for nonzero $\dot{b}$.


\subsection{Natural orbital occupancies} \label{sec:natural_orbital}

The eigenfunctions  $\Phi_j$ of the one-body density matrix are known as the natural orbitals, and
the corresponding eigenvalues $\lambda_j$ are known as the occupancies of the natural orbitals.
\begin{equation}  \label{eq:def_natural_orbitals}
\int dy \, g_{TG}(x,y;t) \Phi_j(y;t) = \lambda_j(t)\,\Phi_j(x;t)\,,\quad j=0,1,2,\ldots
\end{equation}
These occupancies provide a useful characterization of how `condensed' a Bose gas is: when one of
the $\lambda_j$ values is macroscopically dominant, the system is considered to be Bose-condensed in
a single mode \cite{ODLRO}.  The ground-state Tonks-Girardeau gas in a harmonic trap is known to be
quasi-condensed from this perspective because the largest $\lambda_j$ scales as $\sim\sqrt{N}$
instead of as $\sim{N}$ \cite{GirardeauWrightTriscari_PRA2001}.  The dynamics of the natural orbital
occupancies has been widely studied for the TG gas and for the corresponding lattice system as a
measure of the dynamics of the condensate fraction when the system is driven out of equilibrium
\cite{natural_orbital_dynamics_various}.

In our case of trap modulation, if $\Phi_j(x;0)$ is initially a natural orbital, it is an eigenstate
of $g_{TG}(x,y;0)$. Using  Eq.\  \eqref{eq:gt}, one can check 
by substitution into Eq.\  \eqref{eq:def_natural_orbitals} that 
\begin{equation}
\Phi_j(x;t) ~=~ \frac{1}{\sqrt{b}}\; \exp\left[-\tfrac{im}{2\hbar}\tfrac{\dot{b}}{b}x^2\right]\; \Phi_j({x\ov b};0)
\end{equation}
is an eigenstate of $g_{TG}(x,y;t)$ at time $t$, with the \emph{same} eigenvalue
$\lambda_j(t)=\lambda_j(0)$.  In other words, the natural orbital occupancies, and hence the degree
of Bose condensation, stays unchanged during the dynamics, irrespective of whether or not there is a
resonance.  In fact, this result (invariance of $\lambda_j$ for the TG gas in the continuum) is more
general and stays true for \emph{arbitrary} time-dependent variations of the strength of the
harmonic trap, including trap release and trap quenches.

\subsection{Comparison to free fermions} \label{sec:free_fermions}

It is instructive to contrast the behavior of the driven Tonks-Girardeau gas with that of the driven
ideal Fermi gas, onto which it is mapped.

The one-body density matrix for the ideal Fermi gas, $g_{F}(x,y;t)$, is given by an expression
identical to Eq.\ \eqref{eq:TG_densmatr}, except that the sign factors $\mathfrak a$ are absent.

The initial density matrix $g_{F}(x,y;0)$ can be written explicitly as Hankel 
determinants, just like the bosonic $g_{TG}(x,y;0)$ \cite{Forrester2003}.  The time-dependence
$g_{F}(x,y;t)$ is obtained from the initial  $g_{F}(x,y;0)$ by the same scaling transformation as
Eq.\ \eqref{eq:gt}. 

The diagonal part of $g_{F}$ and
$g_{TG}$ are identical for the free Fermi and TG gases, since ${\mathfrak a}(x)^2=1$ in Eq.\ \eqref{eq:TG_densmatr}.   The density profiles are identical not
only in the ground state but also at all later times: $\rho_{F}(x;0)= \rho_{TG}(x;0)$ and
\begin{equation}
\rho_{F}(x;t) = {1\ov b}\rho_{F}\big({x\ov b};0\big) = \rho_{TG}(x;t)
\end{equation}

The momentum distribution profile of the trapped free fermions at equilibrium, $n_{F}(p,0)$, is
well-known to have the same form as the spatial density profile.  The time-dependent $n_{F}(p,t)$ is
obtained using Eq.\ \eqref{eq:mp}, using $g_{F}(x,y;0)$ instead of $g_{TG}(x,y;0)$.  For the free
Fermi gas, the dynamics of the momentum distribution is known 
\cite{MinguzziGangardt2005} to be a rescaling:
\begin{equation}
n_F(p,t) = B(t) \,n_F(B(t) p,0)\,,
\end{equation}
where $B = b/\sqrt{1+b^2 \dot b^2/\om_0^2 }$.  The form of the momentum distribution remains the
same and gets rescaled as time evolves.

\section{Finite small interactions: mean-field treatment} \label{sec:GP}

\begin{figure}[tb]
\centering
\includegraphics[width=\columnwidth]{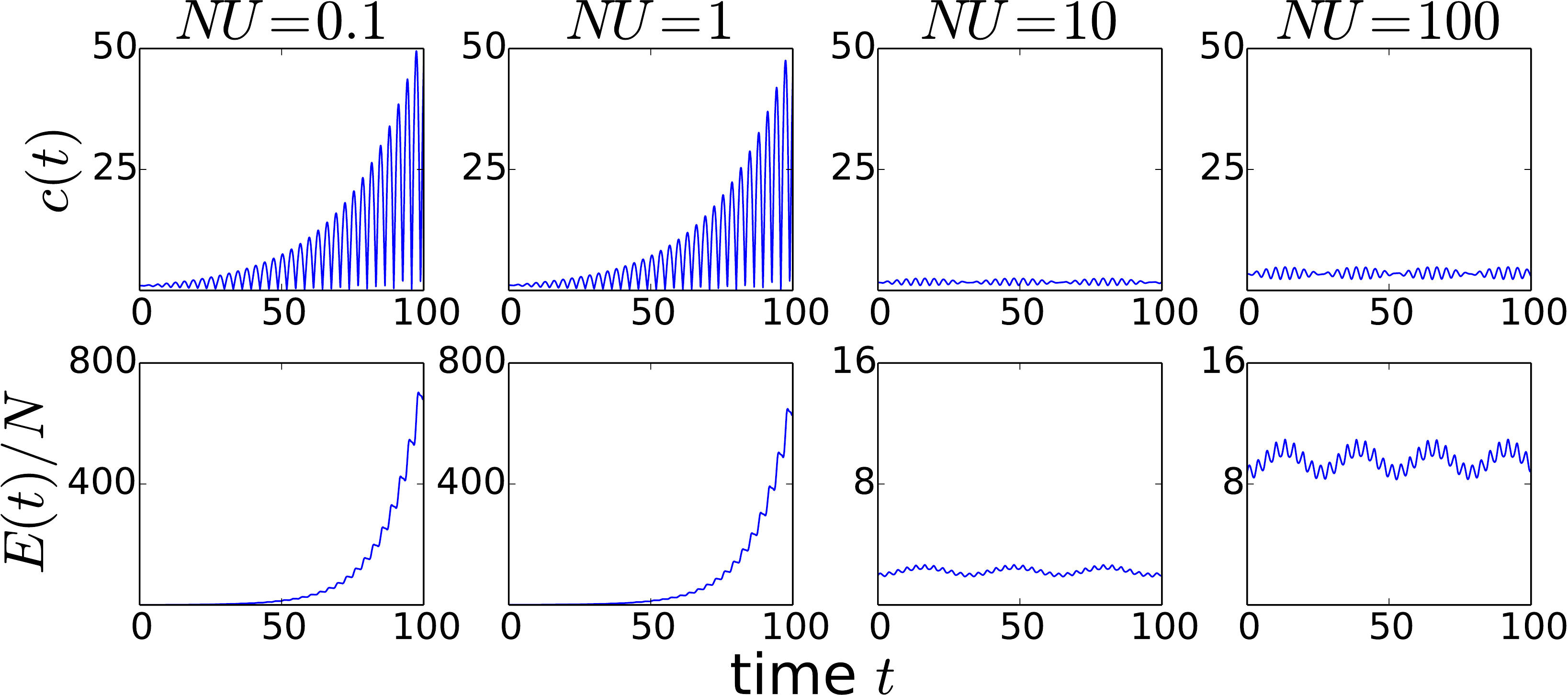}
\caption{\label{fig:cE1} The cloud size $c(t)$ and energy $E(t)$ in the Gross-Pitaevskii
  description, for a range of $NU$ at $\Omega=2\omega_0$, with relatively weak driving
  $\lambda=0.16$.  Exponential resonances are seen for $NU=1$ and smaller, but not for $NU=10$ or
  larger.  Time and energy are expressed in trap units, Eq.\ \eqref{eq:trapunits}.  }
\end{figure}

\begin{figure}[tb]
\centering
\includegraphics[width=\columnwidth]{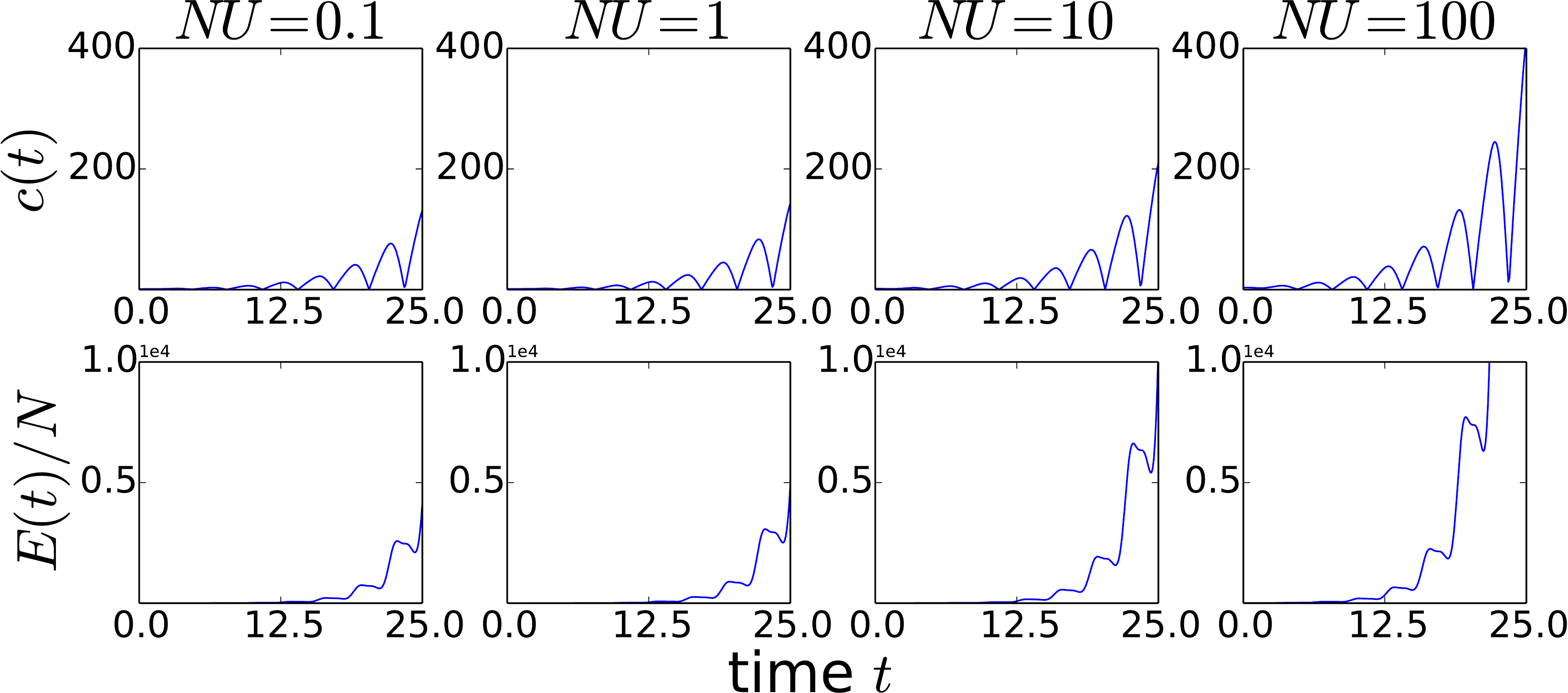}
\caption{\label{fig:cE2}
 The cloud size $c(t)$ and energy $E(t)$ in the Gross-Pitaevskii description, for a range
  of $NU$ at $\Omega=2\omega_0$, with strong driving $\lambda=0.8$.  Exponential resonances appear for all $NU$
  values shown here.  Time and energy are expressed in trap units, Eq.\ \eqref{eq:trapunits}.  }
\end{figure}

Except for the $U=0$ and $U=\infty$ systems, an exact treatment is not possible for the driven
interacting system.  However for small interactions we can use a mean field treatment --- the
Gross-Pitaevskii description --- to study the driving dynamics.  Using a variational ansatz, we
reduce the description of the driving dynamics to the evolution of the size $c(t)$ of the cloud.  The
equation of motion for $c(t)$ in this treatment turns out to be similar to Eq.\ \eqref{eq:b}
governing the scaling parameter $b(t)$ in the exactly solvable cases.

At mean field, the Bose gas is regarded as a quasi-condensate described by the Gross-Pitaevskii
equation \cite{gross-nc20, pitaevskii-jetp13}
\begin{equation}\label{eq:GPE}
i\hb \pDer{}{t}\Psi_{\rm mf} = -{\hb^2\ov 2m}\pDer{^2}{x^2}\Psi_{\rm mf} + {m \om^2(t)  \ov 2} x^2
\Psi_{\rm mf} + U |\Psi_{\rm mf}|^2 \Psi_{\rm mf}\,. 
\end{equation}
Here $\Psi_{\rm mf}(x,t)$ is to be regarded as a condensate `wavefunction' which describes the
$N$-particle system.  We have  normalized $\Psi_{\rm mf}(x,t)$ to $N$.  The mean field
description is best suited to small $U$ and large $N$, and all results in this section should be
interpreted accordingly.

Since driving of the trap strength dominantly excites breathing dynamics, it is appropriate to use a
time-dependent variational ansatz where the cloud size $c(t)$ is a variational parameter:
\begin{equation}\label{eq:vwf}
\Psi_{\rm mf, var} = \sqrt{N \ov c }\left({m \om_0 \ov \pi \hb}\right)^{1/4} \exp\left[-{m \om_0\ov2 \hb}\Big( {x^2\ov  c^2} + i \beta(t) x^2\Big)   \right]\,,
\end{equation}
Optimisation of the variational ansatz constrains the time dependence of $c(t)$ and $\beta(t)$ and
gives equations of motion for these parameters.  This is a standard and widely used technique for
analyzing Gross-Pitaevskii dynamics, dating back to Ref.\
\cite{PerezGarcia_Cirac_Lewenstein_Zoller_variational}.

The imaginary part in the wave function in Eq. \eqref{eq:vwf} is necessary because time evolution starting
from a real wavefunction produces an imaginary component.  However, the two parameters turn out to
be not independent but simply related: $\beta(t) = \tfrac{1}{\om_0}\partial_t{\ln}c(t)$.  There is
thus effectively a single dynamical parameter describing the system, namely the cloud size $c(t)$.
The resulting equation of motion for $c(t)$ is found to be
\begin{equation}\label{eq:c} 
\ddot c +  \omega(t)^2 c = {\om_0^2\ov c^3} + \sqrt{m\om_0^3 \ov 2\pi \hb^3 } {N U \ov c^2} \, ,
\end{equation}
and the energy is found to be 
\begin{equation} \label{eq:variational:energy}
E =N {\hb \om_0 \ov 2} \Big[ {\dot c^2\ov 2 \om_0^2} + {\om(t)^2 c^2 \ov 2 \om_0^2} + {1\ov 2c^2} + \sqrt{m\ov2\pi\hb^3 \om_0} {N U \ov c } \Big]\,.
\end{equation}
In comparison to the exact treatment for $U=0$ in terms of the scaling parameter $b(t)$, the last
term in each of Eqs.\ \eqref{eq:c} and \eqref{eq:variational:energy} provides the effect of
interactions.  We note that the interaction always appears in the combination of $NU$.  This
effective interaction parameter can be large even when we are well within the weak-coupling regime
($U\ll1$) where the Gross-Pitaevskii description is meaningful. 

The initial condition for \eqref{eq:c} is fixed by requiring that the system starts form the ground
state at time $t=0$, and so $\dot c(0)=0$ and $c(0)$ is the unique positive real solution to 
\begin{equation}
c^4(0)- \sqrt{ m\ov 2 \pi\hb^3\om_0} {N U} \,c(0)-1=0 \,.  
\end{equation}

Figures \ref{fig:cE1} and \ref{fig:cE2} show the time evolution of $c$ and $E$ for a range of $NU$
at the primary resonance, with $\lambda=0.16$ and $\lambda=0.8$ respectively. At weaker driving $\lambda=0.16$, the
exponentially resonant behaviour is destroyed when $NU$ increases beyond $\sim5$, while at strong
driving $\lambda=0.8$ the exponential resonances are robust up to large $NU$.  To interpret this result,
we note that the many-body spectrum of interacting bosons in a trap deviates from equal spacing
\cite{QuantumBreathingMode}.  As a result, driving of the same frequency will not connect
successively infinite number of excited eigenstates.  However, if the driving strength $\lambda$ is
not infinitesimal, the individual resonances are broadened, so that the deviations from equal spacing may be
overcome.  This explains why at $\lambda=0.16$ the exponential resonance is still seen up to moderate
$NU$, while at  $\lambda=0.8$ the phenomenon is seen up to even large $NU$.

\begin{figure}[tb]
\centering
\includegraphics[width=\columnwidth]{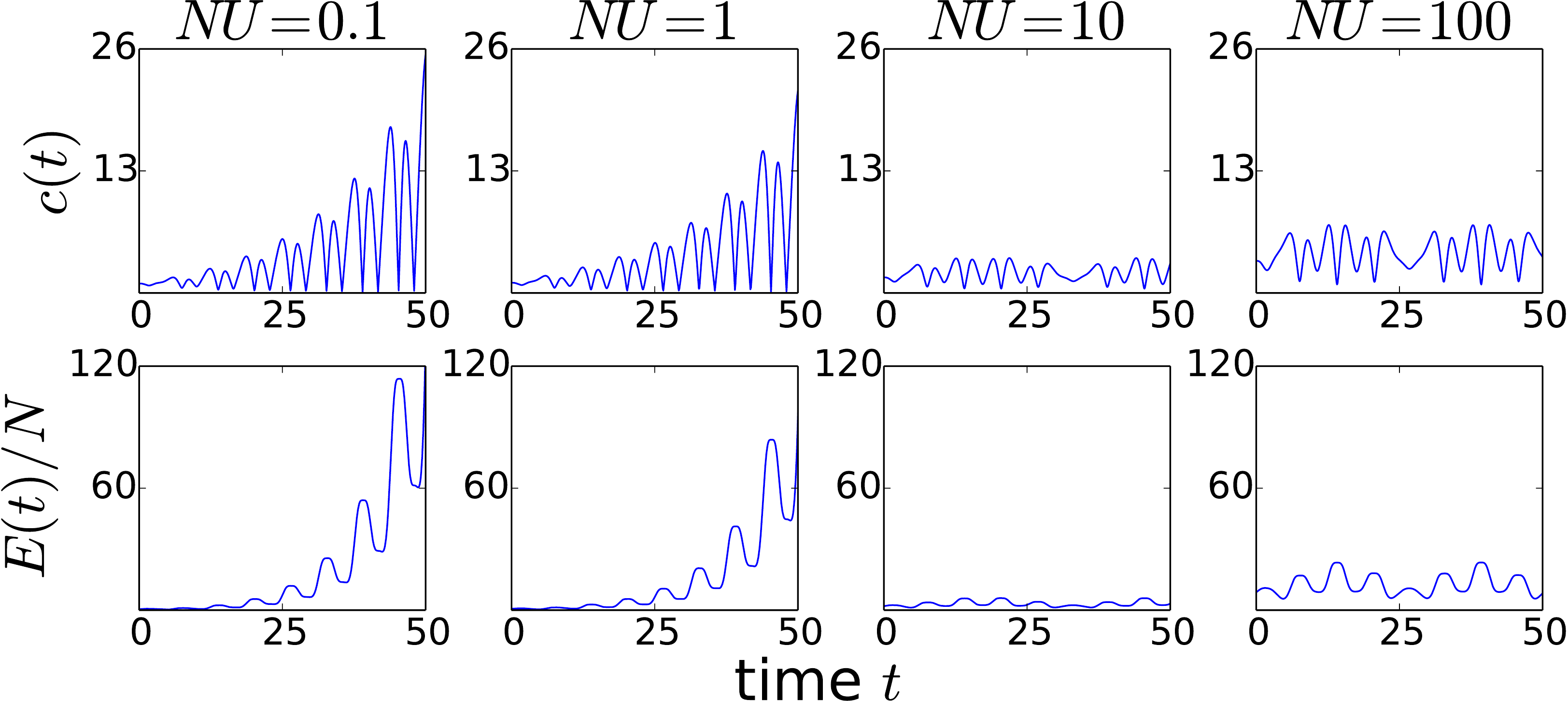}
\caption{
The cloud size $c(t)$ and energy $E(t)$ in the Gross-Pitaevskii description, for a range
  of $NU$ at $\Omega=\omega_0$ (sub-resonance), with strong driving $\lambda=0.8$.  Exponential
  resonance is only seen at moderate values of $NU$. 
Time and energy are expressed in
trap units, Eq.\ \eqref{eq:trapunits}. 
} 
\label{fig:cE3}
\end{figure}

Figure \ref{fig:cE3} shows the breakdown of exponential resonance for increasing $NU$ at the
sub-resonance $\Om=\om_0$, with $\lambda=0.8$.  Becasue the sub-resonances are weaker than the primary
resonance, it may be expected that the exponential resonance does not survive up to large $NU$, even
for strong driving.  The data in Figure \ref{fig:cE3} shows this to be true.

\section{Summary and Discussion} \label{sec:discussion}

Using exact solutions at $U=0$ and $U=\infty$ and a mean-field treatment at finite small $U$, we
have provided a fairly thorough account of resonant behavior of the Lieb-Liniger gas in a perfectly
harmonic trap, under modulations of the trapping frequency.  The solvable limits have resonances at
the same frequencies, and the shift and broadening of the resonances are also identical.  

For the Tonks-Girardeau gas ($U=\infty$), the momentum distribution shows recurrent `fermionization'
but only after resonant driving for some time.  This is similar but not quite identical to what
happens after large quenches of the trap strength \cite{MinguzziGangardt2005}.  We have also
demonstrated that the degree of condensation, as measured by the natural orbital occupancies
$\lambda_i$, stays completely unchanged for the trapped Tonks-Girardeau gas, not only for modulatory
driving but for arbitrary changes of the trapping strength.

Our treatment of the Tonks-Girardeau gas using the scaling transformation is similar in spirit to
Refs.\ \cite{MinguzziGangardt2005, GangardtPustilnik_PRA2008, GritsevDemler_NJP2010,
  Calabrese_TG_traprelease_JSM2013}, who studied trap release and trap quenches for this gas.  Exact
solutions using scaling transformations are also possible in Calogero-Sutherland gases in harmonic
traps; Ref.\ \cite{Sutherland_PRL1998} has also considered periodic driving of such gases.

Using the Gross-Pitaevskii description together with a single-parameter variational description, we
have also treated the regime of small interactions.  This treatment is meaningful for a large number
of particles, $N\gg1$.  Treating a 1D Bose gas as a condensate is well-known to be strongly
approximate.  However, the mean-field treatment provides what we believe to be a satisfactory
qualitative description of the effect of interactions on the resonances.  The exponential growth of
energy and of size oscillations is now seen to happen only for small enough interactions and for
large enough driving.  This is consistent with the fact that the energy eigenstates relevant to
breathing-mode oscillations are no longer equally spaced \cite{QuantumBreathingMode}.  It is
remarkable that the interplay of driving strength and deviation from equal spacing should be
well-described by the mean-field treatment, even though the mean-field description does not a priori
contain information about the many-body eigenspectrum.

The present work opens up various new open questions, of which we list a few below.  

At finite interactions, we have focused on the response of the size, i.e., breathing-mode
oscillations.  This was motivated partly in order to compare with the exact behavior at $U=0$ and
$U=\infty$, where only size oscillations occur, due to the exact scaling form of the solution.
Clearly, at finite $U$ the full response of the cloud will involve much more complex dynamics,
including shape distortions.  The situation is similar to Ref.\ \cite{HaqueZimmer}, where the
dominant dynamics in interaction ramps was extracted through consideration of size dynamics, even
though the full dynamics includes shape distortions, as seen through the time evolution of the
kurtosis of the density profile.

Other than going beyond size dynamics, the description of the finite-$U$ case could also be improved
by going beyond the Gross-Pitaevskii description.  A more accurate hydrodynamic description can be
obtained for general values of $U$ by appealing to the Bethe ansatz solution of the Lieb-Liniger
model \cite{MF_allU, MF_largeU}.  It is expected that the single-parameter description (size
dynamics description) should be reasonable at small $U$, which we have treated, and perhaps also at
large $U$, for which a corresponding formulation remains an open problem.  It is expected that for
intermediate $U$ the size dynamics might not be a very complete description; a full numerical
investigation with the hydrodynamic equations might be worthwhile for this regime.

In a realistic experimental realization, the confining potential will not be perfectly harmonic.
This means the energy levels will not be equally spaced.  Thus, the exponential nature of the
resonances, on which we have focused, will be modified.  Other corrections to the system, such as
corrections to one-dimensionality, corrections to the Lieb-Liniger nature of the interactions, and
finite temperatures, may also disturb the position/nature of the resonances, and might be worth
studying in the context of experiment.  Also, it would be interesting to investigate
higher-dimensional cases, where the mean-field treatment is more reasonable but there is no exact
solvability at $U\to\infty$.  However, for the 2D isotropic trap, the breathing-mode-related
eigenstates are spaced at $2\omega_0$ for any value of the contact interaction
\cite{PitaevskiiRosch_PRA97}.  This suggests exponential resonances at any interaction strength, for
isotropic driving of 2D trapped Bose gases.

\section*{Acknowledgments}

We thank A.~Eckardt for helpful discussions.

\end{document}